\documentclass[aps,prl,twocolumn,superscriptaddress,showpacs,preprintnumbers,amsmath,amssymb]{revtex4-1}
\usepackage{graphicx}
\usepackage{dcolumn}
\usepackage{lineno} 
\usepackage{epstopdf} 
\usepackage{amsmath}
\usepackage{xcolor} 
\RequirePackage{xspace} 
\usepackage{relsize} 
\usepackage{tabularx}    
\usepackage{color}  
 
\def\babar{\mbox{\slshape B\kern-0.1em{\smaller A}\kern-0.1em B\kern-0.1em{\smaller A\kern-0.2em R}}} 
    
\def\Dbar  {\ensuremath{\overline D}\xspace}
\def\Kbar  {\ensuremath{\overline K}\xspace}
\def\ccbar  {\ensuremath{c\overline c}\xspace} 
\def\KS    {\ensuremath{K^0_{\scriptscriptstyle S}}\xspace}

\mathchardef\Upsilon="7107
\def\Y#1S{\ensuremath{\Upsilon{(#1S)}}\xspace}  
\def\cm    {\ensuremath{{\rm \,cm}}\xspace} 
\def\invfb {\ensuremath{\mbox{\,fb}^{-1}}\xspace}
\def\order {{\ensuremath{\cal O}}\xspace}
\def\etal  {{\it et~al.}}
\def\CP {\ensuremath{C\!P}\xspace}
\def\FB {\ensuremath{F\!B}\xspace}
\newcommand{\stat}{\ensuremath{\mathrm{(stat.)}}\xspace}
\newcommand{\syst}{\ensuremath{\mathrm{(syst.)}}\xspace}
\newcommand{\mev}{~\mathrm{\,Me\kern -0.1em V}}
\newcommand{\gev}{~\mathrm{Ge\kern -0.1em V}}
\newcommand{\mevc}{~\mathrm{Me\kern -0.1em V\!/}c}
\newcommand{\gevc}{~\mathrm{Ge\kern -0.1em V\!/}c}
\newcommand{\mevcc}{~\mathrm{\,Me\kern -0.1em V\!/}c^2}
\newcommand{\gevcc}{~\mathrm{Ge\kern -0.1em V\!/}c^2}

\graphicspath{{ps}}

\begin{document}

\preprint{\vbox{ \hbox{ }
                        \hbox{Belle Preprint {\it 2017-06}}
                        \hbox{KEK Preprint {\it 2017-2}}
                        \hbox{arXiv:1712.00619 [hep-ex]} 
}}
 
\title{\quad\\[0.5cm]
Search for {\boldmath $\CP$} violation in the {\boldmath $D^{+}\to\pi^{+}\pi^{0}$} decay at Belle}

%%% Paper:    D+ -> pi+ pi0 CPV
%%% Journal:  Physical Review D (Rapid Communication)
%%% Contacts: V. Babu (varghese@tifr.res.in)
%%%           K. Trabelsi (karim.trabelsi@kek.jp)
%%%           G. Mohanty (gmohanty@tifr.res.in)
%%%           T. Aziz (aziz@tifr.res.in)
%%% Non-responding authors or those who said NO are commented out.
%%% ====================================================================
%%% Click the RELOAD button on your web browser to see the updated file.
%%% ====================================================================
%%% Use \input{author} to insert this material into your latex file.
%%%%% Force institutions to appear in alphabetical order when typeset.
\noaffiliation
%\affiliation{Aligarh Muslim University, Aligarh 202002}
\affiliation{University of the Basque Country UPV/EHU, 48080 Bilbao}
\affiliation{Beihang University, Beijing 100191}
%\affiliation{University of Bonn, 53115 Bonn}
\affiliation{Budker Institute of Nuclear Physics SB RAS, Novosibirsk 630090}
\affiliation{Faculty of Mathematics and Physics, Charles University, 121 16 Prague}
%\affiliation{Chiba University, Chiba 263-8522}
\affiliation{Chonnam National University, Kwangju 660-701}
\affiliation{University of Cincinnati, Cincinnati, Ohio 45221}
\affiliation{Deutsches Elektronen--Synchrotron, 22607 Hamburg}
%\affiliation{University of Florida, Gainesville, Florida 32611}
%\affiliation{Department of Physics, Fu Jen Catholic University, Taipei 24205}
\affiliation{Justus-Liebig-Universit\"at Gie\ss{}en, 35392 Gie\ss{}en}
\affiliation{Gifu University, Gifu 501-1193}
%\affiliation{II. Physikalisches Institut, Georg-August-Universit\"at G\"ottingen, 37073 G\"ottingen}
\affiliation{SOKENDAI (The Graduate University for Advanced Studies), Hayama 240-0193}
%\affiliation{Gyeongsang National University, Chinju 660-701}
\affiliation{Hanyang University, Seoul 133-791}
\affiliation{University of Hawaii, Honolulu, Hawaii 96822}
\affiliation{High Energy Accelerator Research Organization (KEK), Tsukuba 305-0801}
\affiliation{J-PARC Branch, KEK Theory Center, High Energy Accelerator Research Organization (KEK), Tsukuba 305-0801}
%\affiliation{Hiroshima Institute of Technology, Hiroshima 731-5193}
\affiliation{IKERBASQUE, Basque Foundation for Science, 48013 Bilbao}
%\affiliation{University of Illinois at Urbana-Champaign, Urbana, Illinois 61801}
\affiliation{Indian Institute of Science Education and Research Mohali, SAS Nagar, 140306}
\affiliation{Indian Institute of Technology Bhubaneswar, Satya Nagar 751007}
\affiliation{Indian Institute of Technology Guwahati, Assam 781039}
\affiliation{Indian Institute of Technology Hyderabad, Telangana 502285}
\affiliation{Indian Institute of Technology Madras, Chennai 600036}
\affiliation{Indiana University, Bloomington, Indiana 47408}
\affiliation{Institute of High Energy Physics, Chinese Academy of Sciences, Beijing 100049}
\affiliation{Institute of High Energy Physics, Vienna 1050}
\affiliation{Institute for High Energy Physics, Protvino 142281}
%\affiliation{Institute of Mathematical Sciences, Chennai 600113}
\affiliation{University of Mississippi, University, Mississippi 38677}
\affiliation{INFN - Sezione di Napoli, 80126 Napoli}
\affiliation{INFN - Sezione di Torino, 10125 Torino}
\affiliation{Advanced Science Research Center, Japan Atomic Energy Agency, Naka 319-1195}
\affiliation{J. Stefan Institute, 1000 Ljubljana}
\affiliation{Kanagawa University, Yokohama 221-8686}
\affiliation{Institut f\"ur Experimentelle Kernphysik, Karlsruher Institut f\"ur Technologie, 76131 Karlsruhe}
%\affiliation{Kavli Institute for the Physics and Mathematics of the Universe (WPI), University of Tokyo, Kashiwa 277-8583}
\affiliation{Kennesaw State University, Kennesaw, Georgia 30144}
\affiliation{King Abdulaziz City for Science and Technology, Riyadh 11442}
\affiliation{Department of Physics, Faculty of Science, King Abdulaziz University, Jeddah 21589}
\affiliation{Korea Institute of Science and Technology Information, Daejeon 305-806}
\affiliation{Korea University, Seoul 136-713}
\affiliation{Kyoto University, Kyoto 606-8502}
\affiliation{Kyungpook National University, Daegu 702-701}
\affiliation{\'Ecole Polytechnique F\'ed\'erale de Lausanne (EPFL), Lausanne 1015}
\affiliation{P.N. Lebedev Physical Institute of the Russian Academy of Sciences, Moscow 119991}
\affiliation{Faculty of Mathematics and Physics, University of Ljubljana, 1000 Ljubljana}
\affiliation{Ludwig Maximilians University, 80539 Munich}
%\affiliation{Luther College, Decorah, Iowa 52101}
\affiliation{University of Malaya, 50603 Kuala Lumpur}
\affiliation{University of Maribor, 2000 Maribor}
\affiliation{Max-Planck-Institut f\"ur Physik, 80805 M\"unchen}
\affiliation{School of Physics, University of Melbourne, Victoria 3010}
%\affiliation{Middle East Technical University, 06531 Ankara}
\affiliation{University of Miyazaki, Miyazaki 889-2192}
\affiliation{Moscow Physical Engineering Institute, Moscow 115409}
\affiliation{Moscow Institute of Physics and Technology, Moscow Region 141700}
\affiliation{Graduate School of Science, Nagoya University, Nagoya 464-8602}
%\affiliation{Kobayashi-Maskawa Institute, Nagoya University, Nagoya 464-8602}
%\affiliation{Nara University of Education, Nara 630-8528}
%\affiliation{Nara Women's University, Nara 630-8506}
\affiliation{National Central University, Chung-li 32054}
\affiliation{National United University, Miao Li 36003}
\affiliation{Department of Physics, National Taiwan University, Taipei 10617}
\affiliation{H. Niewodniczanski Institute of Nuclear Physics, Krakow 31-342}
\affiliation{Nippon Dental University, Niigata 951-8580}
\affiliation{Niigata University, Niigata 950-2181}
\affiliation{University of Nova Gorica, 5000 Nova Gorica}
\affiliation{Novosibirsk State University, Novosibirsk 630090}
\affiliation{Osaka City University, Osaka 558-8585}
%\affiliation{Osaka University, Osaka 565-0871}
\affiliation{Pacific Northwest National Laboratory, Richland, Washington 99352}
%\affiliation{Panjab University, Chandigarh 160014}
%\affiliation{Peking University, Beijing 100871}
\affiliation{University of Pittsburgh, Pittsburgh, Pennsylvania 15260}
\affiliation{Punjab Agricultural University, Ludhiana 141004}
%\affiliation{Research Center for Electron Photon Science, Tohoku University, Sendai 980-8578}
%\affiliation{Research Center for Nuclear Physics, Osaka University, Osaka 567-0047}
\affiliation{Theoretical Research Division, Nishina Center, RIKEN, Saitama 351-0198}
%\affiliation{RIKEN BNL Research Center, Upton, New York 11973}
%\affiliation{Saga University, Saga 840-8502}
\affiliation{University of Science and Technology of China, Hefei 230026}
%\affiliation{Seoul National University, Seoul 151-742}
%\affiliation{Shinshu University, Nagano 390-8621}
\affiliation{Showa Pharmaceutical University, Tokyo 194-8543}
\affiliation{Soongsil University, Seoul 156-743}
\affiliation{University of South Carolina, Columbia, South Carolina 29208}
\affiliation{Stefan Meyer Institute for Subatomic Physics, Vienna 1090}
\affiliation{Sungkyunkwan University, Suwon 440-746}
\affiliation{School of Physics, University of Sydney, New South Wales 2006}
\affiliation{Department of Physics, Faculty of Science, University of Tabuk, Tabuk 71451}
\affiliation{Tata Institute of Fundamental Research, Mumbai 400005}
\affiliation{Excellence Cluster Universe, Technische Universit\"at M\"unchen, 85748 Garching}
\affiliation{Department of Physics, Technische Universit\"at M\"unchen, 85748 Garching}
\affiliation{Toho University, Funabashi 274-8510}
%\affiliation{Tohoku Gakuin University, Tagajo 985-8537}
\affiliation{Department of Physics, Tohoku University, Sendai 980-8578}
\affiliation{Earthquake Research Institute, University of Tokyo, Tokyo 113-0032}
%\affiliation{Department of Physics, University of Tokyo, Tokyo 113-0033}
\affiliation{Tokyo Institute of Technology, Tokyo 152-8550}
\affiliation{Tokyo Metropolitan University, Tokyo 192-0397}
%\affiliation{Tokyo University of Agriculture and Technology, Tokyo 184-8588}
\affiliation{University of Torino, 10124 Torino}
%\affiliation{Utkal University, Bhubaneswar 751004}
\affiliation{Virginia Polytechnic Institute and State University, Blacksburg, Virginia 24061}
\affiliation{Wayne State University, Detroit, Michigan 48202}
\affiliation{Yamagata University, Yamagata 990-8560}
\affiliation{Yonsei University, Seoul 120-749}
 \author{V.~Babu}\affiliation{Tata Institute of Fundamental Research, Mumbai 400005} % Tata
 \author{K.~Trabelsi}\affiliation{High Energy Accelerator Research Organization (KEK), Tsukuba 305-0801}\affiliation{SOKENDAI (The Graduate University for Advanced Studies), Hayama 240-0193} % KEK
 \author{G.~B.~Mohanty}\affiliation{Tata Institute of Fundamental Research, Mumbai 400005} % Tata
 \author{T.~Aziz}\affiliation{Tata Institute of Fundamental Research, Mumbai 400005} % Tata
 \author{D.~Greenwald}\affiliation{Department of Physics, Technische Universit\"at M\"unchen, 85748 Garching} % TUM
% \author{A.~Abdesselam}\affiliation{Department of Physics, Faculty of Science, University of Tabuk, Tabuk 71451} % Tabuk
 \author{I.~Adachi}\affiliation{High Energy Accelerator Research Organization (KEK), Tsukuba 305-0801}\affiliation{SOKENDAI (The Graduate University for Advanced Studies), Hayama 240-0193} % KEK
% \author{K.~Adamczyk}\affiliation{H. Niewodniczanski Institute of Nuclear Physics, Krakow 31-342} % Krakow
% \author{J.~K.~Ahn}\affiliation{Korea University, Seoul 136-713} % Korea
 \author{H.~Aihara}\affiliation{Department of Physics, University of Tokyo, Tokyo 113-0033} % Tokyo
 \author{S.~Al~Said}\affiliation{Department of Physics, Faculty of Science, University of Tabuk, Tabuk 71451}\affiliation{Department of Physics, Faculty of Science, King Abdulaziz University, Jeddah 21589} % Tabuk
% \author{K.~Arinstein}\affiliation{Budker Institute of Nuclear Physics SB RAS, Novosibirsk 630090}\affiliation{Novosibirsk State University, Novosibirsk 630090} % BINP
% \author{Y.~Arita}\affiliation{Graduate School of Science, Nagoya University, Nagoya 464-8602} % Nagoya
 \author{D.~M.~Asner}\affiliation{Pacific Northwest National Laboratory, Richland, Washington 99352} % PNNL
 \author{H.~Atmacan}\affiliation{University of South Carolina, Columbia, South Carolina 29208} % SouthCarolina
% \author{V.~Aulchenko}\affiliation{Budker Institute of Nuclear Physics SB RAS, Novosibirsk 630090}\affiliation{Novosibirsk State University, Novosibirsk 630090} % BINP
% \author{T.~Aushev}\affiliation{Moscow Institute of Physics and Technology, Moscow Region 141700} % MIPT
% ...
% Ljubljana
% \author{N.~Zwahlen}\affiliation{\'Ecole Polytechnique F\'ed\'erale de Lausanne (EPFL), Lausanne 1015} % Lausanne
\collaboration{Belle Collaboration}

\begin{abstract}
We search for $\CP$ violation in the charged charm meson decay $D^{+}\to\pi^{+}\pi^{0}$, based on a data sample corresponding to an integrated luminosity of $921\invfb$ collected by the Belle experiment at the KEKB $e^{+}e^{-}$ asymmetric-energy collider.
The measured $\CP$-violating asymmetry is $[+2.31\pm1.24\stat\pm0.23\syst]\%$, which is consistent with
the standard model prediction and has a significantly improved precision compared to previous results.
\end{abstract}


\maketitle
{\renewcommand{\thefootnote}{\fnsymbol{footnote}}}
\setcounter{footnote}{0}

%%%%%%%%%%%%%%%%%%%%%%%%%%%%%%%%%%%%%%%%%%%%%%%%%%%%%%%%%%%%%%%%%%
Within the standard model (SM), the violation of charge-parity ($\CP$) symmetry in the charm system is expected to be small
[$\order(10^{-3})$] owing to suppression from the Glashow-Iliopoulos-Maiani mechanism~\cite{GIM}. 
These order-of-magnitude estimates~\cite{Gino} suffer from large uncertainties~\cite{Brod} due to nonperturbative long-distance
effects resulting from a finite charm-quark mass.
The problem came to the fore in 2012, when the world average of the difference in $\CP$-violating asymmetries between
$D^{0}\to K^{+}K^{-}$ and $D^{0}\to\pi^{+}\pi^{-}$ decays was measured to be $\Delta A_{\CP}=(-0.656\pm 0.154)\%$~\cite{Amhis};
here, each asymmetry is
\begin{equation}
 A_{\CP}(D\to f) =  \frac{\Gamma(D\to f) - \Gamma(\Dbar\to\overline{f})}{\Gamma(D\to f) + \Gamma(\Dbar\to\overline{f})},
\end{equation}
where $\Gamma(D\to f)$ and $\Gamma(\Dbar\to\overline{f})$ are the decay rates for a given process and its $\CP$ conjugate, respectively.
This led to much discussion as to whether the result was consistent with the SM or a signature of new physics (NP). 
Though the current $\Delta A_{\CP}$ value is consistent with zero~\cite{Amhis-new}, it is important to study those decay channels
expected by the SM to exhibit negligible $\CP$ violation.

Singly Cabibbo-suppressed decays like $D^{+}\to \pi^{+}\pi^{0}$~\cite{charge} are excellent candidates to probe $\CP$ violation in the charm sector~\cite{Grossman}.
Such decays require additional strong and weak phases besides those in the tree diagram to have a sizable $\CP$ asymmetry.
The phases can appear in either a strong or an electroweak loop (e.g., box diagram).
As the former produces only isospin singlets, it cannot contribute to the $I=2$ final state of $\pi^{+}\pi^{0}$.
On the other hand, electroweak loop diagrams have too small an amplitude of $\order(10^{-6})$ for the interference to manifest $\CP$ violation.
Any $\CP$ asymmetry found in these channels would therefore point to NP~\cite{Grossman, Buccella}.
In particular, the authors of Ref.~\cite{Grossman} suggested looking for $\CP$ violation in $D^{+}\to\pi^{+}\pi^{0}$ as well as 
verifying a sum rule that relates individual asymmetries of the three isospin-related $D\to\pi\pi$ decays as potential
NP probes. The sum rule, which reduces the theoretical uncertainty due to strong interaction effects, can be characterized by the ratio
\begin{equation} 
R=
\frac{\left|{\cal A}_1\right|^2-\left|\bar{{\cal A}}_1\right|^2+\left|{\cal A}_2\right|^2-\left|\bar{{\cal A}}_2\right|^2-\frac{2}{3}(\left|{\cal A}_3\right|^2-\left|\bar{{\cal A}}_3\right|^2)} 
{\left|{\cal A}_1\right|^2+\left|\bar{{\cal A}}_1\right|^2+\left|{\cal A}_2\right|^2+\left|\bar{{\cal A}}_2\right|^2+\frac{2}{3}(\left|{\cal A}_3\right|^2+\left|\bar{{\cal A}}_3\right|^2)},
\end{equation} 
where ${\cal A}_{1}$, ${\cal A}_{2}$, and ${\cal A}_{3}$ are the amplitudes of  $D^0\to\pi^+\pi^-$, $D^0\to\pi^0\pi^0$, and $D^+\to\pi^+\pi^0$, respectively;
$\bar{{\cal A}}_{1}$, $\bar{{\cal A}}_{2}$, and $\bar{{\cal A}}_{3}$ are those of their $\CP$ conjugates.
The amplitudes are normalized so that
\begin{equation}        
\left|{\cal A}_k\right|^2\propto\frac{{\cal B}_{k}}{\tau_{0\,(+)}\,p_{k}},
\end{equation}
where ${\cal B}_{k}$ is the branching fraction of the decay $D\to\pi_{i}\pi_{j}$,
$\tau_{0\,(+)}$ is the appropriate $D^{0}\,(D^{+})$ lifetime, and
\begin{equation}        
p_{k}=\frac{\{[m^2_D-(m_i+m_j)^2][m^2_D+(m_i-m_j)^2]\}^{\frac{1}{2}}}{2\,m_D},
\end{equation}
is the breakup momentum in the $D$ rest frame.
The indices $i$ and $j$ correspond to the pion daughters.
As the masses of the charged and neutral species of the $D$ or $\pi$ mesons
are close to each other, we consider all $p_{k}$ values to be equal.
We use Eqs.~(3)--(4) and the relation
\begin{equation}
\left|{\cal A}_k\right|^2-\left|\bar{{\cal A}}_k\right|^2=A_{\CP}\left(\left|{\cal A}_k\right|^2+\left|\bar{{\cal A}}_k\right|^2\right)
\end{equation}
to rewrite Eq.\,(2) as
\begin{align}
R =\frac{A_{\CP}(D^0\to\pi^+\pi^-)}{1+\frac{\tau_{D^{0}}}{{\cal B}_{1}}\left(\frac{{\cal B}_{2}}{\tau_{D^{0}}}+\frac{2}{3}\frac{{\cal B}_{3}}{\tau_{D^{+}}}\right)}
+\,\frac{A_{\CP}(D^0\to\pi^0\pi^0)}{1+\frac{\tau_{D^{0}}}{{\cal B}_{2}}\left(\frac{{\cal B}_{1}}{\tau_{D^{0}}}+\frac{2}{3}\frac{{\cal B}_{3}}{\tau_{D^{+}}}\right)} \\\nonumber
-\,\frac{A_{\CP}(D^+\to\pi^+\pi^0)}{1+\frac{3}{2}\frac{\tau_{D^{+}}}{{\cal B}_{3}}\left(\frac{{\cal B}_{2}}{\tau_{D^{0}}}+\frac{{\cal B}_{1}}{\tau_{D^{0}}}\right)}.
\end{align}
If the value of $R$ is consistent with zero while the $\CP$ asymmetry in $D^{+}\to\pi^{+}\pi^{0}$ is nonzero~\cite{Grossman},
it would be an NP signature.

A test of the above sum rule requires the measurement of the time-integrated $\CP$ asymmetries $A_{\CP}(D^{0}\to\pi^{+}\pi^{-})$,
$A_{\CP}(D^{0}\to\pi^{0}\pi^{0})$, and $A_{\CP}(D^{+}\to\pi^{+}\pi^{0})$.
The current world average of $A_{\CP}(D^{0}\to\pi^{+}\pi^{-})$ is $(+0.13\pm 0.14)\%$~\cite{pdg}.
Three years ago, Belle measured $A_{\CP}(D^{0}\to\pi^{0}\pi^{0})$ as $[-0.03\pm 0.64\stat\pm 0.10\syst]\%$~\cite{Nisar_P}.
However the charged-mode asymmetry measured by CLEO has an uncertainty of $2.9\%$~\cite{mendez} and therefore limits
the precision with which the above sum rule can be tested.

We present herein an improved measurement of $\CP$ asymmetry for the channel $D^{+} \to\pi^{+}\pi^{0}$
using the full $e^{+}e^{-}$ collision data sample recorded by the Belle experiment~\cite{Abashian} at the KEKB
asymmetric-energy collider~\cite{Kurokawa}. The data sample was recorded at three different center-of-mass (CM) energies:
at the $\Y4S$ and $\Y5S$ resonances and $60\mev$ below the $\Y4S$ peak, with corresponding integrated luminosities of $711$,
$121$ and $89\invfb$, respectively.
The detector components relevant for the study are a tracking system comprising a silicon vertex detector  and a 50-layer central drift chamber
(CDC), a particle identification device that consists of a barrel-like arrangement of time-of-flight scintillation counters (TOF) and an array of
aerogel threshold Cherenkov counters (ACC), and a CsI(Tl) crystal electromagnetic calorimeter (ECL). All these components are located
inside a superconducting solenoid that provides a 1.5\,T magnetic field.

For the measurement, we consider an exclusive sample of $D^{\pm}$ mesons tagged by $D^{*\pm}\to D^{\pm}\pi^{0}$ decays, and another that is not tagged by the $D^{*\pm}$ decays.
The former sample has a better signal-to-noise ratio while the latter has more events. For optimal sensitivity, we combine their asymmetry measurements.

From a simultaneous fit to the invariant-mass ($M_{D}$) distributions of the $\pi^{\pm}\pi^{0}$ samples,
we determine the raw asymmetry
 \begin{equation}
A_{\rm raw}^{\pi\pi} = \frac{N(D^{+}\to\pi^{+} \pi^{0})-N(D^{-}\to\pi^{-} \pi^{0})}{N(D^{+}\to\pi^{+}\pi^{0})+N(D^{-}\to\pi^{-} \pi^{0})},
\end{equation}
where $N(D^{+}\to\pi^{+}\pi^{0})$ and $N(D^{-}\to\pi^{-} \pi^{0})$ are the yields for the signal and its $\CP$-conjugate process, respectively.
$A_{\rm raw}^{\pi\pi}$  has three contributing terms:
\begin{equation}
A_{\rm raw}^{\pi\pi} = A_{\CP}^{\pi\pi} + A_{\FB} + A_{\varepsilon}^{\pi^{\pm}}.
\end{equation}
The first term, $A_{\CP}^{\pi\pi}$, is the true asymmetry.
The forward-backward asymmetry, $A_{\FB}$, arises due to interference between the amplitudes mediated by a virtual photon, a $Z^{0}$
boson, and higher-order effects~\cite{FB-paper1,FB-paper2,FB-paper3} in $e^{+}e^{-} \to\ccbar$.
It is an odd function of the cosine of the $D^{*\pm}$ polar angle, $\theta^{*}$, in the CM frame.
The pion-detection efficiency asymmetry, $A_{\varepsilon}^{\pi^{\pm}}$, is a function of the $\pi^{\pm}$ momentum and polar angle.

We make use of the high-statistics normalization channel $D^{+}\to\KS\pi^{+}$ to correct the measured asymmetry for $A_{\FB}$ and
$A_{\varepsilon}^{\pi^{\pm}}$.
As both signal and normalization decays arise from the same underlying process, $A_{\FB}$ should be identical for them.
This assumption has been verified by checking the consistency of the $\cos\theta^{*}$ distribution between the two decays.
Further, we expect $A_{\varepsilon}^{\pi^{\pm}}$ to be the same if the two channels have similar pion momentum and polar-angle distributions.
The angle distributions for the two channels are found to be identical.
Though there is a small difference between the momentum distributions, it has been verified to have a negligible impact on the measurement.
The raw asymmetry for the normalization channel is thus
 \begin{equation}
A_{\rm raw}^{K\pi} = A_{\CP}^{K\pi} + A_{\FB} + A_{\varepsilon}^{\pi^{\pm}}, 
\end{equation}
where $A_{\CP}^{K\pi}$ is the $\CP$ asymmetry of $D^{+}\to\KS\pi^{+}$; this has been measured to
be $[-0.363\pm 0.094\stat\pm 0.067\syst]\%$~\cite{brko}, including the $\CP$ asymmetry induced by $K^{0}$-$\Kbar^0$
mixing and the difference in interactions of $K^{0}$ and $\Kbar^{0}$ mesons with the detector material.
The difference in the raw asymmetries is
\begin{equation}
\Delta A_{\rm raw} \equiv A_{\rm raw}^{\pi \pi} -  A_{\rm raw}^{K \pi}   =  A_{\CP}^{\pi \pi} - A_{\CP}^{K \pi},
\end{equation}
which leads to 
\begin{equation}
\label{imp_eqn}
 A_{\CP}^{\pi\pi} = A_{\CP}^{K\pi} + \Delta A_{\rm raw}. 
\end{equation}  

Monte Carlo (MC) simulated events are used to devise and optimize the selection criteria;
the size of the MC sample corresponds to an integrated luminosity six times that of the data.
We perform the optimization by maximizing the signal significance, $N_{\rm sig}/\sqrt{N_{\rm sig}+N_{\rm bkg}}$, where $N_{\rm sig}$ $(N_{\rm bkg})$
is the number of signal (background) events expected within a $\pm 3\sigma$ window ($\sigma = 15.3\mevcc$)  around the nominal $D$ mass~\cite{pdg}.
The branching fraction of the signal channel used in the $N_{\rm sig}$ calculation is the current world average, $1.24\times 10^{-3}$~\cite{pdg}.
The background level is corrected for a possible data-MC difference by comparing yields in the $M_D$ sidebands of $1.70$--$1.76$  and $1.92$--$2.00\gevcc$.
 
Charged-track candidates must originate from near the $e^{+}e^{-}$ interaction point (IP), with an impact parameter
along the $z$ axis and in the transverse plane of less than $3.0$ and $1.0\cm$, respectively.
(The $z$ axis is the direction opposite the $e^{+}$ beam.)
They must have a momentum greater than $840\mevc$.
They are treated as pions if the likelihood ratio, ${{\cal L}_{\pi}}/({{\cal L}_{\pi}}+{{\cal L}_K})$, is greater than 0.6, where ${\cal L}_{\pi}$ and
${\cal L}_{K}$ are the pion and kaon likelihoods, respectively. These are calculated with information from the CDC, TOF and ACC.
This requirement, when applied to charged particles with a momentum distribution similar to that of the signal decay, yields a pion
identification efficiency of approximately $88\%$ and a kaon-to-pion misidentification probability of about $7\%$.

The high-momentum (``hard'') $\pi^{0}$ candidates that would originate from two-body $D$ decay are reconstructed from pairs of
photons by requiring the diphoton invariant mass to be within $\pm 16\mevcc$ of the nominal $\pi^{0}$ mass~\cite{pdg}.
The hard $\pi^{0}$ daughter photons in the barrel, forward-- and backward--endcap regions of the ECL are required to have an energy
greater than 50, 100 and $150\mev$, respectively.
(The barrel, forward-- and backward--endcap regions span the polar angle ranges $32.2$--$128.0^{\circ}$, $12.4$--$31.4^{\circ}$ and
$130.7$--$155.1^{\circ}$, respectively.)
The thresholds for the endcap photons are higher due to the higher beam background. 
The hard $\pi^{0}$ must have a momentum greater than $1.06\gevc$. 

Charged $D$ meson candidates are formed by combining a charged-pion with a hard-$\pi^{0}$ candidate, and requiring the
resultant $M_{D}$ distribution to lie within $\pm200\mevcc$ of the nominal $D$ mass~\cite{pdg}.
For $D^{*+}$ reconstruction in the tagged sample, low-momentum (``soft'') $\pi^{0}$ candidates are reconstructed from a pair of photon candidates
whose energy criteria are optimized  for each ECL region;
the corresponding  values are listed in Table~\ref{phot_cuts}.
The soft-$\pi^{0}$ invariant mass is required to be within an optimized window, $125$--$143\mevcc$.
It is verified during optimization that the $\pi^{0}$ mass distributions in simulations are in agreement with control data
consisting of a high-statistics sample of $D^{+}\to K^{-}\pi^{+}\pi^{+}$ decays, with the $D^+$ arising from $D^{*+}\to D^{+}\pi^{0}$.

\begin{table}[tbh]
\caption{Optimized requirements on the soft-$\pi^{0}$ photon energies (ECL region) in the tagged sample. }
\label{phot_cuts}
\begin{ruledtabular}
\begin{tabular}{cll}
Case  &  $E_{\gamma 1}$ criterion &  $E_{\gamma 2}$ criterion \\
\hline 
1 	& 	$>$\,46 MeV (barrel) & $>$\,46 MeV (barrel)\\
2 	& 	$>$\,36 MeV (barrel) & $>$\,68 MeV (forward endcap)\\
3 	& 	$>$\,30 MeV (barrel) & $>$\,44  MeV (backward endcap)\\
\end{tabular} 
\end{ruledtabular}
\end{table}  

For the tagged sample, $D^{*}$ candidates are formed by combining  $D$ mesons with soft $\pi^{0}$ candidates such that the mass difference
between the $D^{*}$ and $D$ candidates, $\Delta M$, lies within an optimized  window of $139$--$142\mevcc$.
This corresponds approximately to a $\pm 1.5 \sigma$ signal region, where $\sigma$ is the $\Delta M$ resolution.
For the fit to extract $A_{\CP}$ (described below), two intervals of $D^{*}$ CM momentum with
different signal-to-background ratios are chosen: $p^{*}_{D^{*}} > 2.95\gevc$ and $2.50\gevc< p^{*}_{D^{*}} <2.95\gevc$.
The first corresponds to an optimized $p^{*}_{D^{*}}$ criterion with maximal signal significance.
The second interval is added to increase the statistical sensitivity of the measurement, while ensuring that the lower bound
excludes $D^{*}$ mesons from a $B$-meson decay, as the latter might introduce a nontrivial $\CP$ asymmetry.
 
After the above selection criteria are applied, we find that about $3\%$ of events have multiple $D^{*}$ candidates.
We perform a best-candidate  selection (BCS) to remove  spurious $D^{*}$ candidates
formed from fake soft-$\pi^{0}$ mesons.
This is done by retaining, for each event, the candidate whose $\Delta M$ value lies closest to the mean of
the $\Delta M$ distribution, $140.69\mevcc$.
For events with multiple $D^{*}$ candidates, with at least one of them being the true candidate, the
BCS successfully identifies the correct one around $65\%$ of the time.
As the spurious $D^*$ candidates also correspond to  true $D$ candidates, this component peaks in the $M_{D}$
distribution.
By performing the BCS, we ensure that only one $D$ candidate is selected per event, and so avoid overestimating
the signal component in the $M_{D}$ fits.

If there are no suitable $D^{*}$ candidates found in an event, the charged $D$ candidates, if any, are considered for the untagged sample.
Here, we require that the $D$ CM momentum be above an optimized threshold of $2.65\gevc$.
In case there are multiple $D$ candidates in the event, the one with the daughter $\pi^0$ candidate having a 
reconstructed mass closest to the nominal $\pi^0$ mass~\cite{pdg} is chosen.
If there are still multiple surviving candidates, the one whose charged-pion daughter has the smallest transverse impact parameter is retained.
About $2\%$ of events in the untagged sample have multiple $D$ candidates;
for such events, with at least one of them being the true candidate,
the BCS successfully identifies the correct one around $66\%$ of the time.

For the normalization channel, we reconstruct  $K^{0}_{S}$ candidates from pairs
of oppositely charged tracks that have an invariant mass within $30\mevcc$ $(\pm 5 \sigma)$ of the nominal $\KS$
mass. The transverse impact parameter of the track candidates  is required to be larger than $0.02\cm$
for high-momentum ($>1.5\gevc$) and $0.03\cm$ for low-momentum ($<1.5\gevc$) $\KS$ candidates.
The $\pi^{+}\pi^{-}$ vertex must be displaced from the IP by a minimum (maximum) transverse (longitudinal)
distance of $0.22\cm$ ($2.40\cm$) for high-momentum candidates and $0.08\cm$ ($1.80\cm$) for the remaining
candidates.
The direction of the $\KS$ momentum must be within 0.03\,rad (0.10\,rad) of the direction between the IP
and the vertex for high-momentum (remaining) candidates.
The surviving $\KS$ candidates  are kinematically constrained to their nominal masses~\cite{pdg}.
Candidate events for the $D^{+}\to\KS\pi^{+}$ channel are selected with  essentially the same requirements as for signal,
except that we require the $D$ candidate mass to lie within $\pm80\mevcc$ of the nominal $D$ mass;
the tighter criterion is due to the better mass resolution with an all-charged final state.
Similar to the signal channel described earlier, nonoverlapping tagged and untagged samples are formed.

\begin{figure}[htb]  
\begin{center}
 \begin{tabular}{cc}
   \includegraphics[width=0.5\textwidth]{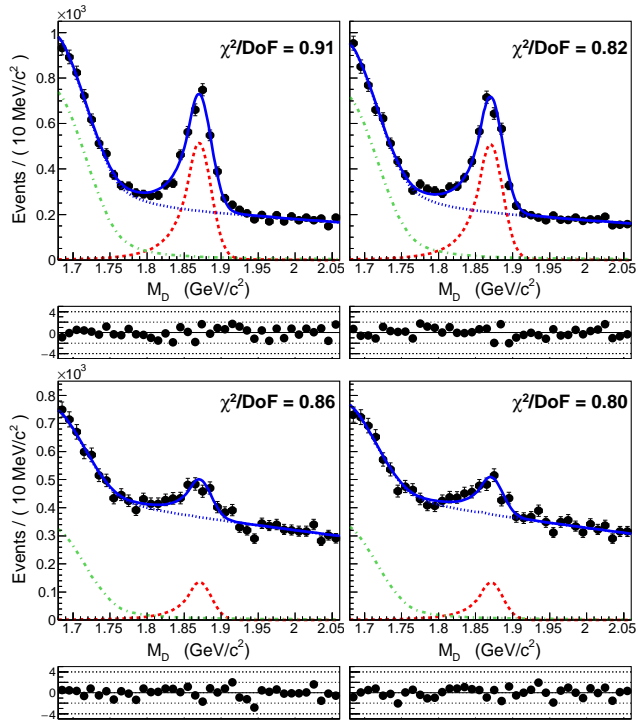}
\end{tabular}
\caption{Invariant mass distributions for the $\pi^{\pm}\pi^{0}$ system for the tagged $D\to\pi\pi$ sample in the intervals $p^{*}_{D^{*}}> 2.95\gevc$ (top) and $2.50\gevc < p^{*}_{D^{*}} < 2.95\gevc$ (bottom).
Left (right) panels correspond to $D^{+}$ ($D^{-}$) samples.
Points with error bars are the data. The solid blue curves are the results of the fit.
The red dashed, blue dotted and green dash-dotted curves show the signal, total- and peaking-background contributions, respectively.
The normalized residuals are shown below each distribution, and the post-fit $\chi^{2}$ per degree of freedom ($\chi^{2}$/d.o.f.) is given in each panel.}
\label{fit_pipi}   
\end{center}
\end{figure}

\begin{figure}[htb]  
\begin{center}
 \begin{tabular}{cc} 
  \includegraphics[width=0.5\textwidth]{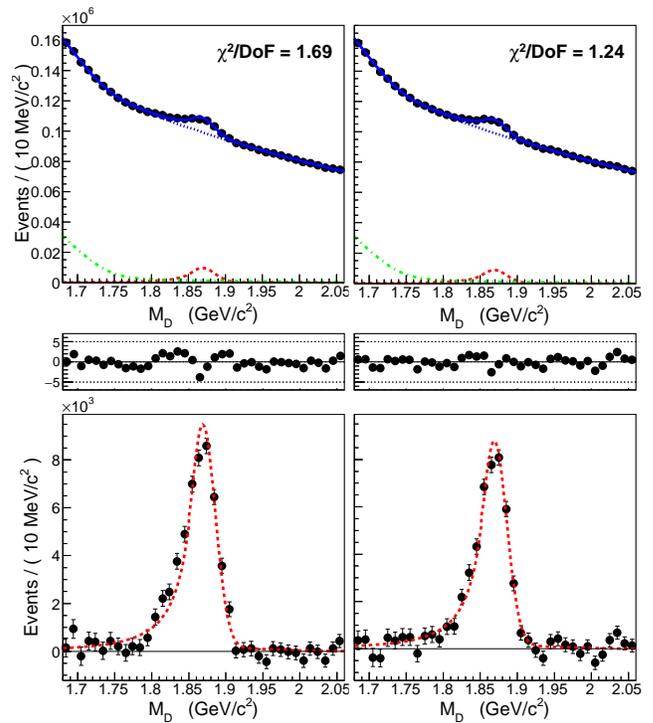}
\end{tabular}
\caption{Invariant mass distributions for the  $\pi^{\pm}\pi^{0}$ system for the untagged $D\to\pi\pi$ sample.
The top two panels are the full distributions with signal and background components, while the bottom two show the
corresponding background-subtracted distributions. Left (right) panels correspond to $D^{+}$ ($D^{-}$) samples.
Points with error bars, colored curves, and residual plots are described in the caption of Fig.~\ref{fit_pipi}.}
\label{fit_pipi_untag} 
\end{center} 
\end{figure}

A fitting range of $1.68$--$2.06\gevcc$ in $M_{D}$ is chosen for the signal $D\to\pi\pi$ channel.
For the tagged sample, a simultaneous unbinned maximum-likelihood fit of the two $p^{*}_{D^{*}}$ intervals
and oppositely charged $D$ meson candidates is performed.
Similarly, for the untagged sample, a simultaneous binned maximum-likelihood  fit of oppositely charged
$D$ meson candidates is done.
We use a combination of a Crystal Ball (CB)~\cite{Skwarnicki} and a Gaussian function to model the
signal peak for both tagged and untagged fits.
The background in the tagged fit is parametrized by the sum of a reversed CB and a linear polynomial,
while that for the untagged fit uses a quadratic rather than a linear polynomial.
All signal shape parameters for the tagged fit are fixed to MC values except for an overall mean and a
width scaling factor, which are floated. 
We introduce the scaling factor to account for the possible difference between data and simulations.
For the untagged fit, all shape parameters are fixed to MC values, aside from the overall mean, which is
floated, and the width scaling factor, which is fixed from the tagged-data fit.
For the background, the cutoff and tail parameters of the reversed CB are fixed from MC events, and all other shape parameters are floated.
For the tagged fit, the two $p^{*}_{D^{*}}$ intervals are required to have a common signal asymmetry but have separate background asymmetries. 
For the tagged sample, the total signal yield obtained from the fit is $6632\pm 256$ with $A^{\pi\pi}_{\mathrm{raw}}=(+0.52\pm1.92)\%$;
the corresponding results for the untagged sample are $100934 \pm 1952$ and $(+3.77\pm1.60)\%$.
The quoted uncertainties are statistical. 
Figures~\ref{fit_pipi} and \ref{fit_pipi_untag} show the projections of the simultaneous fit performed on the tagged and untagged data samples, respectively.

%%%%%%%%%%%%%%%%************%%%%%%%%%%%%%%%%%%%%%%%%*******************

\begin{figure}[htb]    
\begin{center}
 \begin{tabular}{cc}
   \includegraphics[width=0.5\textwidth]{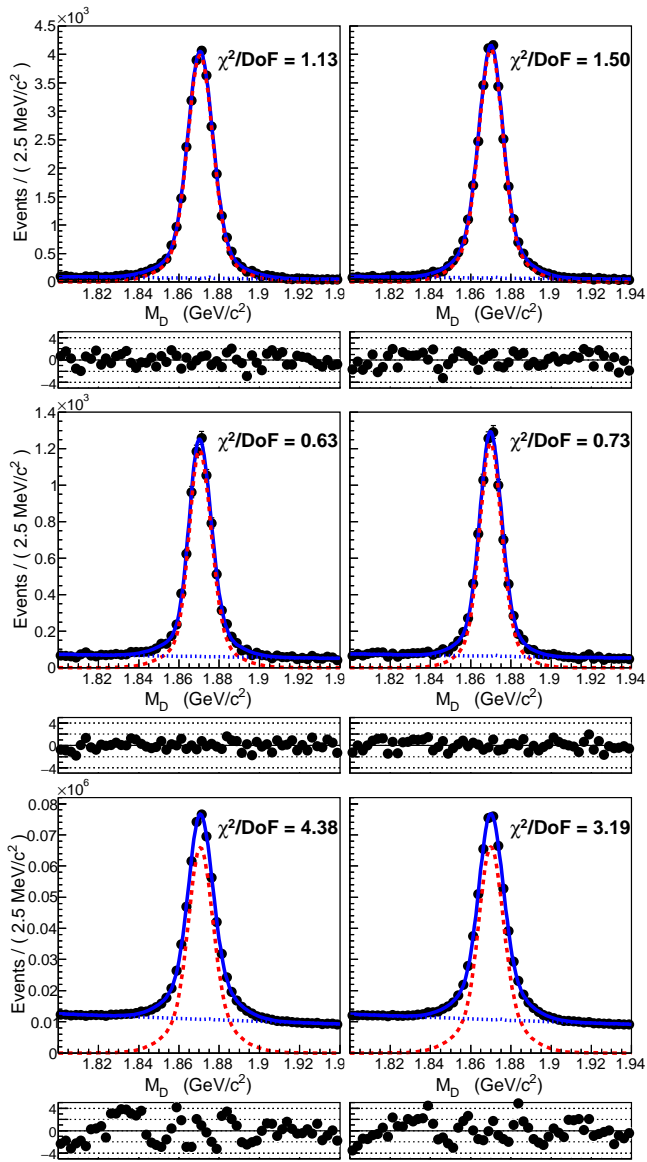}
\end{tabular}
\caption{Invariant mass distributions for the $\KS\pi^{\pm}$ system for the normalization channel, $D\to\KS\pi^{\pm}$, in the intervals  $p^{*}_{D^{*}}> 2.95\gevc$ (top) and $2.50\gevc < p^{*}_{D^{*}} < 2.95\gevc$ (middle) for the tagged sample, and for the untagged sample (bottom).
Left (right) panels correspond to $D^+$($D^-$) samples.
Points with error bars, colored curves, and residual plots are described in the caption of Fig.~\ref{fit_pipi}. }
\label{fit_kspi}  
\end{center}
\end{figure}

For the $D^+\to\KS\pi^+$ normalization channel, a fitting range of $1.80$--$1.94\gevcc$ is chosen and the simultaneous fits for the tagged sample,
with two $p^{*}_{D^{*}}$ intervals, and the untagged sample are performed as for the $D\to\pi\pi$ signal channel.
The narrower fitting range can be afforded because of the better $D$-mass resolution.
The signal peak is modeled with the sum of a Gaussian and an asymmetric Gaussian function, with all shape parameters floated.
The background shape is parametrized with a linear polynomial, whose slope is floated.
The total signal yield obtained from the tagged fit is $68434\pm 308$  with $A_{\mathrm{raw}}^{K\pi}= (-0.29\pm 0.44)\%$;
the corresponding results for the untagged sample are $982029 \pm 1797$ and $(-0.25\pm0.17)\%$.
The quoted uncertainties are again statistical.
Figure~\ref{fit_kspi} shows the projections of the simultaneous fit performed on the tagged and untagged data samples.

From the results of the fit to the signal and normalization channels, we calculate $\Delta A_{\rm raw}$ (tagged)
$=(+0.81\pm1.97\pm0.19)\%$ and $\Delta A_{\rm raw}$ (untagged) $ =(+4.02 \pm 1.61 \pm 0.32) \% $.
The first uncertainty quoted in each measurement is statistical and the second is systematic (see below).
A combination of the two~\cite{err-add} gives
\begin{equation}
\Delta A_{\rm raw}=(+2.67 \pm 1.24 \pm 0.20)\%,
\end{equation}
which, in conjunction with the world average of $A_{\CP}(D^+ \to \KS \pi^+)$~\cite{pdg}, results in
\begin{equation}
A_{\CP}(D^+\to\pi^+\pi^0)=(+2.31\pm1.24\pm0.23)\%.
\end{equation}    

The major sources of systematic uncertainty for the $A_{\CP}$ measurement are: (i) uncertainty in the
signal  and background shapes  for the $D\to\pi\pi $ fits, (ii)  uncertainty in modeling the
peaking-background shape, and (iii) uncertainty in the $A_{\CP}$ measurement for the normalization channel.
Source (i) arises from fixing some of the shape parameters to MC values.
Its contribution to the systematic uncertainties is estimated by constructing an ensemble of fits, randomizing the
fixed parameters with Gaussian distributions whose mean and width are set to MC values
and then extracting the RMS of the $A_{\rm raw}$ distribution obtained from the fits.
The peaking background of source (ii) is due to misreconstructed $D$ or $D_{s}$ meson decays and exhibits a
broad peaking structure shifted to the left of the signal peak (Figs.~\ref{fit_pipi} and \ref{fit_pipi_untag}).
As it is only partially present in the fitting range, the reversed-CB shape is subject to uncertainty.
We vary the lower $M_{D}$ threshold between 1.68 to $1.72\gevcc$ in steps of $10\mevcc$ and
then refit to assess the impact on the signal's $A_{CP}$ determination.
For source (iii), we rely on the world average of $A_{\CP}(D^+\to\KS\pi^+)$~\cite{pdg}.
The various sources of systematic uncertainties and their values are listed in Table~\ref{syst}.
The total uncertainty is  $\pm 0.23\%$.    

\begin{table}[tbh]  
\caption{Summary of systematic uncertainties (\%) on $A_{\CP}$. }
\label{syst}
\begin{ruledtabular}
\begin{tabular}{lccc}
Source  &  $D \to\pi\pi$ & & $D \to\pi\pi$ \\
             &  tagged          & & untagged \\
\hline  
Signal shape 	& 	$\pm0.02$	& & $\pm 0.23$\\ 
Peaking background shape &  $\pm 0.19$ & & $\pm 0.22$\\
\hline 
$\Delta A_{\rm raw}$ measurement &  $\pm 0.19$ & & $\pm 0.32$\\ 
\hline 
$A_{\CP}(D\to\KS\pi)$ measurement & & $\pm 0.12$ &\\
\hline 
Total & & & \\
(combined $A_{\CP}$ measurement) & &  $\pm 0.23$  &\\  
\end{tabular} 
\end{ruledtabular}     
\end{table}

In summary, we have measured the $\CP$-violating asymmetry $A_{\CP}$ for the $D^{+}\to\pi^{+}\pi^{0}$ decay
using $921\invfb$ of data, with the combined result from two disjoint samples: one tagged by the decay $D^{*+}\to D^{+}\pi^{0}$
and the other untagged.
After correcting for the forward-backward asymmetry and detector-induced efficiency
asymmetry, based on the normalization channel $D^{+}\to\KS\pi^{+}$, we obtain
$A_{\CP}(D^{+}\to\pi^{+}\pi^{0})=  [+2.31\pm1.24\stat\pm0.23\syst]\%$.
The result is consistent with the SM expectation of null asymmetry and improves the precision by more than a factor of $2$ over the previous measurement~\cite{mendez}.
Inserting this result into Eq.\,(6) along with the current world averages of $A_{\CP}$ and ${\cal B}$ for $D^0\to\pi^+\pi^-$~\cite{pdg} and
$D^0\to\pi^0\pi^0$~\cite{Nisar_P} decays, as well as $\tau_{0\,(+)}$~\cite{pdg}, we obtain
$R = (-2.2\pm 2.7)\times 10^{-3}$.
The isospin sum rule holds to a precision of three per mille, putting constraints on the NP parameter space~\cite{Grossman}.
As the statistical error of $A_{\CP}(D^0\to\pi^0\pi^0)$, as well as of our result, dominate the total uncertainty on $R$,
we expect a substantial improvement in testing the sum rule from the upcoming Belle II experiment~\cite{belle2}.

We thank the KEKB group for the excellent operation of the
accelerator; the KEK cryogenics group for the efficient
operation of the solenoid; and the KEK computer group,
the National Institute of Informatics, and the 
PNNL/EMSL computing group for valuable computing
and SINET5 network support.  We acknowledge support from
the Ministry of Education, Culture, Sports, Science, and
Technology (MEXT) of Japan, the Japan Society for the 
Promotion of Science (JSPS), and the Tau-Lepton Physics 
Research Center of Nagoya University; 
the Australian Research Council;
Austrian Science Fund under Grant No.~P 26794-N20;
the National Natural Science Foundation of China under Contracts 
No.~10575109, No.~10775142, No.~10875115, No.~11175187, No.~11475187, 
No.~11521505 and No.~11575017;
the Chinese Academy of Science Center for Excellence in Particle Physics; 
the Ministry of Education, Youth and Sports of the Czech
Republic under Contract No.~LTT17020;
the Carl Zeiss Foundation, the Deutsche Forschungsgemeinschaft, the
Excellence Cluster Universe, and the VolkswagenStiftung;
the Department of Science and Technology of India; 
the Istituto Nazionale di Fisica Nucleare of Italy; 
National Research Foundation (NRF) of Korea Grants No.~2014R1A2A2A01005286, No.~2015R1A2A2A01003280,
No.~2015H1A2A1033649, No.~2016R1D1A1B01010135, No.~2016K1A3A7A09005603, No.~2016R1D1A1B02012900; Radiation Science Research Institute, Foreign Large-size Research Facility Application Supporting project and the Global Science Experimental Data Hub Center of the Korea Institute of Science and Technology Information;
the Polish Ministry of Science and Higher Education and 
the National Science Center;
the Ministry of Education and Science of the Russian Federation and
the Russian Foundation for Basic Research;
the Slovenian Research Agency;
Ikerbasque, Basque Foundation for Science and
MINECO (Juan de la Cierva), Spain;
the Swiss National Science Foundation; 
the Ministry of Education and the Ministry of Science and Technology of Taiwan;
and the U.S.\ Department of Energy and the National Science Foundation.


\begin{thebibliography}{30}
\bibitem{charge}
 Unless stated otherwise, the inclusion of charge-conjugate reactions are implied.
\end{thebibliography}
\end{document}